%% file: main.tex
\documentclass[conference]{IEEEtran}
\IEEEoverridecommandlockouts
\usepackage{cite}
\usepackage{amsmath,amssymb,amsfonts}
\usepackage{algorithmic}
\usepackage{graphicx}
\usepackage{textcomp}
\usepackage{xcolor}
\usepackage{booktabs}
\usepackage{multirow}
\usepackage{tabularx}
\usepackage{subcaption}
\usepackage{xspace}
\usepackage{footnote}
\usepackage{color,soul}
\usepackage{comment}
\def\BibTeX{{\rm B\kern-.05em{\sc i\kern-.025em b}\kern-.08em
    T\kern-.1667em\lower.7ex\hbox{E}\kern-.125emX}}
\begin{document}

\title{ReProbe: An Architecture for Reconfigurable\\and Adaptive Probes
\thanks{
This work has been partially supported by the Centro Nazionale HPC, Big Data e Quantum Computing (PNRR CN1 spoke 9 Digital Society \& Smart Cities); the Engineered MachinE Learning-intensive IoT systems (EMELIOT) national research project, which has been funded by the MUR under the PRIN 2020 program (Contract 2020W3A5FY); and the COmmunity-Based Organized Littering (COBOL) national research project, which has been funded by the MUR under the PRIN 2022 PNRR program (Contract P20224K9EK).
}
}

\makeatletter
\newcommand{\linebreakand}{%
  \end{@IEEEauthorhalign}
  \hfill\mbox{}\par
  \mbox{}\hfill\begin{@IEEEauthorhalign}
}
\makeatother

\newcommand{\approach}{ReProbe\xspace}
\newcommand{\etal}{\xspace\textit{et al.}\xspace}

\author{
\IEEEauthorblockN{
Federico Alessi\IEEEauthorrefmark{1},
Alessandro Tundo\IEEEauthorrefmark{2}\IEEEauthorrefmark{1},
Marco Mobilio\IEEEauthorrefmark{1},
Oliviero Riganelli\IEEEauthorrefmark{1},
Leonardo Mariani\IEEEauthorrefmark{1}}
\linebreakand\\
\IEEEauthorblockA{\IEEEauthorrefmark{1}\textit{University of Milano-Bicocca}\\
Milan, Italy \\
f.alessi@campus.unimib.it,
\{name.surname\}@unimib.it}
\and\\
\IEEEauthorblockA{\IEEEauthorrefmark{2}\textit{Vienna University of Technology}\\
Vienna, Austria \\
\{name.surname\}@tuwien.ac.at}
}

\maketitle

\begin{abstract}
Modern distributed systems are highly dynamic and scalable, requiring monitoring solutions that can adapt to rapid changes. Monitoring systems that rely on external probes can only achieve adaptation through expensive operations such as deployment, undeployment, and reconfiguration.

This poster paper introduces ReProbes, a class of adaptive monitoring probes that can handle rapid changes in data collection strategies. ReProbe offers controllable and configurable self-adaptive capabilities for data transmission, collection, and analysis methods. The resulting architecture can effectively enhance probe adaptability when qualitatively compared to state-of-the-art monitoring solutions.
\end{abstract}


\input{sections/01-introduction}

\input{sections/02-approach}
\input{sections/03-evaluation}
\input{sections/04-related-work}

\input{sections/05-conclusions}

\bibliography{references.bib}
\bibliographystyle{IEEEtran}

\end{document}

%% file: sections/01-introduction.tex
%!TeX root = main.tex
 
\section{Introduction}
\label{sec:introduction}
The growing usage of %applications and services built with distributed architectures, such as, 
microservice-based applications, Systems-of-Systems (SoS), and Internet-of-Things (IoT) services make effective and highly-dynamic monitoring of paramount importance to ensure seamless operations in the face of growing complexity~\cite{noor2019framework}. Despite applications can be enhanced with monitoring capabilities through code instrumentation techniques (e.g., Kieker Framework~\cite{Kieker2020}, OpenTelemetry~\cite{opentelemetry}, Prometheus client libraries~\cite{prometheusclientlibs}), this is not always possible as in the case of legacy or third-party systems, where monitoring activities have to be performed by employing external probes\footnote{A probe is a software component responsible for collecting the raw data, such as sampling the CPU consumption of a service or recording the temperature in a room from a sensor.}.

% As cloud systems embrace the architecture of microservices, providing modular, scalable, and isolated functionalities, effective monitoring becomes of paramount importance to ensure seamless operations in the face of growing complexity~\cite{noor2019framework}. While microservices enhance flexibility and efficiency, the continuous and rapid evolution of modern cloud systems pose new challenges to monitoring systems~\cite{abderrahim2017holistic,taherizadeh2018monitoring,usman2022survey}. In fact, state-of-the-art monitoring systems may require to redeploy or reconfigure probes\footnote{A probe is a software component responsible for collecting the raw data, such as sampling the CPU consumption of a service or recording the temperature in a room from a sensor.} to accommodate any change in the data collection strategies~\cite{tundo2023automated}.
State-of-the-art monitoring systems using external probes can be adapted to changes by only performing \emph{expensive} probe-level operations, such as redeploying or reconfiguring probes, to accommodate any change in the data collection strategies~\cite{tundo2023automated}. For instance, changing the set of collected indicators or changing the sampling rate of data collection
% in Metricbeat~\cite{metricbeat} require updating the configuration files and restarting the daemon, or configuring live reloading to dynamically reload configuration files when there are changes. Similarly, when
using Prometheus~\cite{prometheus} and its exporters~\cite{prometheusexporters} requires: (i) updating the Prometheus configuration to change the sampling rate (i.e., scraping interval), (ii) updating the exporter configuration to change the collected indicators, and (iii) at least restarting the exporter when Prometheus is properly configured for hot reload via HTTP API request.

% This short paper \emph{addresses the challenge of flexible data collection by presenting early research results about the design of adaptive monitoring probes} that can quickly change (i) configuration settings (i.e., the sampling rate, the observed target, and the collected indicators), (ii) which ingestion services send data to, (iii) the indicator collection logic, and (iv) the data analysis logic, without requiring redeployment. The class of adaptive probes proposed in this paper, that we called \emph{ReProbe}s, augments probing systems with efficient \emph{probe-level self-adaptive behaviors}, that is, adaptive behaviors local to the components responsible for data collection, in contrast to solutions designed to facilitate probe (un)deployment~\cite{tundo2023automated}. 

This poster paper \emph{presents a hierarchical architecture that can be used to design reconfigurable and adaptive monitoring probes} that can \emph{quickly} change (i) configuration settings (i.e., the sampling rate, the observed target, and the collected indicators), (ii) the ingestion services, (iii) the data collection logic, and (iv) the data analysis logic, without requiring any expensive probe redeployment. The class of adaptive probes presented in this paper, which we called \emph{ReProbe}s, can augment probing systems with efficient \emph{probe-level self-adaptive behaviors}, that is, adaptive behaviors local to the components responsible for data collection, in contrast to solutions designed to facilitate probe (un)deployment~\cite{tundo2023automated}. The proposed architecture exploits the plug-in architectural pattern~\cite{birsan2005plug,richards2015software}  to achieve the desired level of flexibility. Our initial qualitative evaluation shows that \approach can effectively cope with dynamic monitoring scenarios by offering flexible adaptivity.

%% file: sections/02-approach.tex
\newcommand{\pubmanager}{Publishers Manager\xspace}
\newcommand{\colmanager}{Collectors Manager\xspace}
\newcommand{\datamanager}{Data Manager\xspace}
\newcommand{\api}{API\xspace}

\section{\approach}
\label{sec:approach}

\begin{figure}[ht]
    \centering
    \includegraphics[width=0.9\linewidth]{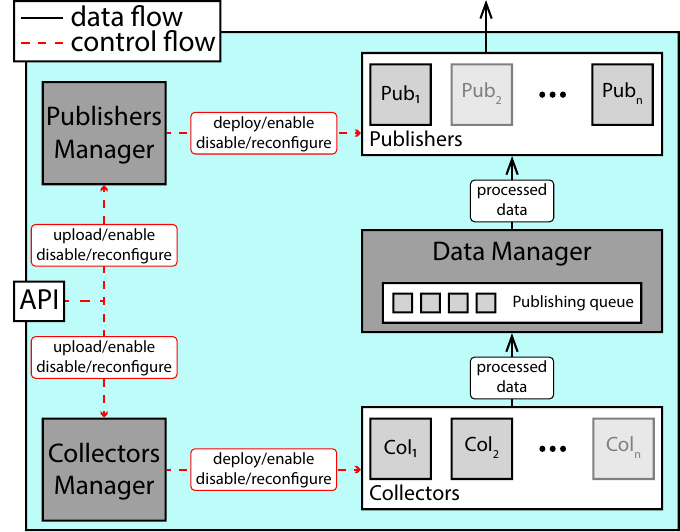}
    \caption{\approach architecture.}
    \label{fig:reprobes}
\end{figure}

%In this section, we present the architecture of \approach that we designed around the \emph{plugin architectural pattern}~\cite{birsan2005plug,richards2015software}.
Figure~\ref{fig:reprobes} shows the main modules of \approach, namely the \emph{\pubmanager}, its collection of \textit{Publishers} (Pub), the \emph{\colmanager}, its collection of \textit{Collectors} (Col), and the \emph{\datamanager}.
Note that \approach defines the architecture of individual self-adaptive probes, which can be used in the context of any monitoring system. In fact, \approach does not impose any restriction on the type of collected data, which can be transmitted through any platform (e.g., Apache Kafka streams~\cite{kafkastreams}) and stored in any time series database (e.g., Elasticsearch~\cite{elasticsearch}). \approach can be used to define probes that embed self-adaptive behaviors that influence \emph{what} and \emph{how} data is collected, and then transmitted outside the probe. 

The \colmanager and \pubmanager are the components responsible for handling the life-cycle of the plugin available in the collection of collectors and publishers, respectively. When an authorized client requests changes through the API interface, the managers add and remove plugins from the collections. Moreover, they also handle the run-time operation of the plugin, namely \emph{instantiating} (i.e., starting a plugin), \emph{destroying} (i.e., eliminating a previously created instance of a plugin), and \emph{reconfiguring} (i.e., update the configuration of a running plugin) them.

A \textit{Col} in the set of the collectors has the responsibility to collect data from the monitored resource and publish the collected data to the \datamanager. The architecture supports the presence of multiple Cols since each Col might be in charge of collecting a different set of indicators. For instance, a Col can be responsible for the collection of CPU utilization indicators (e.g., idle, I/O wait, user, system, and total percentages) while another Col can be responsible for the collection of network-related indicators (e.g., I/O received packets, I/O dropped packets, and I/O bytes). %The two plugins jointly may implement a probe for the collection of indicators about the infrastructure.
Conversely, a publisher has the responsibility of reading data from the \datamanager and publishing the data to ingestion services. Again, the architecture supports the presence of multiple publishers able to deal with different ingestion services. %(e.g., a publisher for Elasticsearch~\cite{elasticsearch} and a different publisher for TimescaleDB~\cite{timescale}). 

Multiple collectors and publishers can be simultaneously active. The data collected by the running collectors is forwarded to the data manager and consumed by the active publishers that send the data to specific external services. For example, let us consider an instance of \approach configured to monitor a load balancer (e.g., HAProxy). It may collect system-level indicators by means of two Cols (e.g., CPU utilization and memory consumption), and application-level indicators by means of another Col (e.g., HAProxy performance statistics). Simultaneously, all the collected observations are sent to Elasticsearch~\cite{elasticsearch} for offline processing. At the same time, application-level indicators are also sent to Kafka streams~\cite{kafkastreams} for an online processing pipeline used to tune HAProxy configuration.

Any external actor (e.g., operator or other services) can interact with \approach through the \api to \emph{upload} new plugin, and to control the life-cycle of the uploaded plugins.

Figure~\ref{fig:reprobes} shows how \approach can reconfigure itself, responding to requests received by the \api, through the red dashed lines, which represent the control flow, and the black lines, which represent the data-flow links.
Note that the adaptations triggered through the \api are purely \emph{reactive}, since the trigger is external to the probe. %Indeed, the choice of the kind of collected data (i.e., the choice of the collector that must be installed and instantiated) and the choice of the services where the data should be published (i.e., the choice of the publisher that must be installed and instantiated) are not choices that the probe can take autonomously.However, each collector is designed to embed self-adaptive behaviors that are autonomously activated by the probe, which can follow both \emph{reactive} and \emph{proactive} adaptations. We describe the collector in details in the next section.

%\section{Collector} \label{sec:collector}
\smallskip 

\newcommand{\controller}{Controller\xspace}

\begin{figure}[ht]
	\centering	\includegraphics[width=0.9\linewidth]{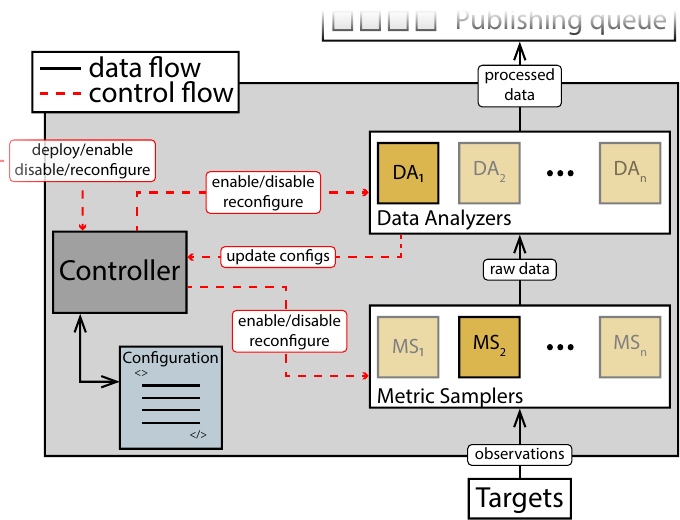}
	\caption{Collector architecture.}
	\label{fig:collectorplugin}
\end{figure}

The \textit{Collector} is the component responsible for collecting a given set of indicators from a target. In the \approach architecture, it embeds self-adaptive behaviors that can be used to proactively adjust the monitoring strategy according to observations.
Figure~\ref{fig:collectorplugin} shows the architecture of a \textit{Collector}, which includes a \textit{\controller}, a set of \textit{Metric Samplers} (MSs), a set of \textit{Data Analyzers} (DAs), and a \emph{configuration}.

The \textit{Metric Sampler} is the component responsible for gathering a given set of indicators from the monitored resource. For instance, it can be the component that reads the CPU or the memory utilization according to a given sampling rate. Each MS inside the plugin encodes a slightly different way of sampling the same data %(e.g., sampling CPU utilization values regularly rather than sampling sparsely as long as values are outside certain boundaries).
(e.g., sampling the bandwidth %passive or active 
measurements based on the recent indicator values to increase accuracy or reduce the overhead). 
%Only one MS can be active at any given time inside a collector.

The \textit{Data Anlyzer} receives the sampled observations from the active MS. It is responsible for any analysis and manipulation (e.g., filtering data, changing its format, and aggregating data) before sending the processed data outside the boundary of the plugin to the \datamanager. 

Only a single MS and DA can be active at any given time.

The \textit{Configuration} stores the information needed by a collector to operate, such as the parameters that influence the behavior of the running DA and MS, as well as the configuration of the target resource.  

The \textit{\controller} is the component in charge of concretely updating the configurations and triggers any reconfiguration of the plugin, if needed.

In addition to processing the collected data, DAs also include the logic to trigger the self-adaptation through the reconfiguration of the metric samplers. For example, if the active DA reports that a specific indicator is stable, it may ask the \controller to update the configuration and reduce the sampling rate of the active MS to save resources (e.g. less overhead on the monitored target or less network traffic). On the other hand, if a given metric is unstable, the same DA may increase the sampling rate. The presence of multiple DAs and multiple MSs allows to collect the observations in different ways and self-adapt the monitoring strategy according to different policies.

Note how each Col is a self-adaptive module itself, running two flows. The first flow is represented by the data gathered from the target by the MS, which flows through the DA and finally outside to the Data Manager of \approach. The second flow is the self-adaptive loop triggered by the DA through the Controller every time active MS needs to be reconfigured. 

\begin{comment}
This architecture can be used to obtain, and also flexibly switch between, multiple types of probes. A \emph{basic probe for memory monitoring} can be obtained by using a single publisher, for instance publishing the collected data on Kafka, and a simple collector that collects memory occupation from the configured target and uses a plain data forwarder as data analysis component. A more \emph{sophisticated non-adaptive memory monitoring probe} can be obtained by running multiple publishers to publish data over multiple streams plus a collector with both a MS that gathers memory occupation and a DA component that computes additional statistics while making data available in the publishing queue. Finally \emph{a self-adaptive probe for memory consumption} can be obtained again running one or more publishers and a MS that gathers memory occupation data, passing the data to a DA component that adjusts the sampling rate depending how far the target is from saturating the available memory (the less memory is available, the higher the sampling rate is). Finally, note that the API of the probe can be exploited to completely reconfigure a running \approach instance, potentially switching from a basic probe for memory monitoring to a self-adaptive probe for memory consumption without interruption.
\end{comment}

%% file: sections/03-evaluation.tex
 %!TeX root = main.tex
 
\section{Qualitative Evaluation}
\label{sec:evaluation}

To assess the proposed solution, we qualitatively compared \approach to both research and commercial monitoring solutions along two dimensions, namely, \emph{features} and \emph{usage scenarios}. The feature-based evaluation compares the adaptation capabilities of the probes available in the considered solutions. The scenario-based evaluation assesses the effectiveness of the probes in the context of two realistic monitoring scenarios.

For the comparison, we selected six representative monitoring solutions for the cloud: two research monitoring tools with adaptive behaviors, JCatascopia~\cite{trihinas2014jacatascopia} and Adaptive FogMon~\cite{colombo2022towards}; and four popular commercial tools~\cite{verginadis2023review}, Prometheus~\cite{prometheus}, Datadog~\cite{datadog}, Zabbix~\cite{zabbix}, and ElasticStack~\cite{elasticstack}.

\subsection{Feature-Based Comparison}

\begin{table*}[!ht]
\centering
\caption{Feature-Based Comparison}
\label{tab:feature-based-comparison}
\resizebox{\textwidth}{!}{%
\begin{tabular}{@{}lccccccc@{}}
\toprule
\multicolumn{1}{c}{\multirow{2}{*}{\textbf{Feature}}} &
  \multicolumn{7}{c}{\textbf{Probe Technologies}} \\ \cmidrule(l){2-8} 
\multicolumn{1}{c}{} &
  \textit{\approach} &
  \begin{tabular}[c]{@{}c@{}}\textit{JCatascopia}\\\textit{Agent}~\cite{trihinas2014jacatascopia}\end{tabular} &
  \begin{tabular}[c]{@{}c@{}}\textit{Adaptive FogMon}\\\textit{Follower}~\cite{colombo2022towards}\end{tabular} &
  \begin{tabular}[c]{@{}c@{}}\textit{ElasticStack}\\\textit{Beats}~\cite{beats}\end{tabular} &
  \begin{tabular}[c]{@{}c@{}}\textit{Prometheus}\\\textit{Exporter}~\cite{prometheusexporters}\end{tabular} &
  \begin{tabular}[c]{@{}c@{}}\textit{Datadog}\\\textit{Agent}~\cite{datadog}\end{tabular} &
  \begin{tabular}[c]{@{}c@{}}\textit{Zabbix}\\\textit{Agent}~\cite{zabbix}\end{tabular} \\ \cmidrule(r){1-1}
Zero-downtime deployment of new collection logic & yes & yes & no & no  & no & no & yes \\
Zero-downtime deployment of new data publishing logic & yes & no & no & no  & no & no & no \\

\hline

API-enabled configuration                & yes & yes & no & no & no & yes & yes \\
Self-adaptive collection logic              & yes & no & partially & no & no & no & no \\
Within-probe data analysis                   & yes & yes & yes & no & no & no & no \\

\hline

Supporting multiple data ingestion services & yes & no & no & yes & no & no & no \\
Simultaneous data publishing      & yes & no & no & no  & no & no & no \\ \bottomrule
\end{tabular}%
}
\end{table*}

We identified seven main adaptive features that are desirable for self-adaptive probes (as summarized in Table~\ref{tab:feature-based-comparison}). Two of these features concern with the deployment of new behaviors, three concern with run-time adaptation, and two concern with the flexibility of the probe. Regarding the extensibility of the probes, we have \emph{zero-downtime deployment of new collection logic} and \emph{zero-downtime deployment of new data publishing logic} features that concern with the capability of adding new ways of collecting data (e.g., to support new indicators or to run new analyses on the collected indicators) and publishing data (e.g., to publish data on a new, previously unsupported, external service). %It is often important that the deployment of new code does not imply stopping the probe, thus missing observations. 
Regarding changes to run-time behaviors, \emph{API-enable configuration} is the possibility to reconfigure a probe from the outside through an API; \emph{self-adaptive collection logic} is the capability to self-adapt the configuration of the data collection logic (e.g., to adaptively modify the sampling rate or logic); and \emph{within-probe data analysis} is the capability to locally analyze data and immediately react to any detected pattern (e.g., changing the data collection logic or reporting an anomaly). Regarding the flexibility of the probe, \emph{supporting multiple data ingestion services} concerns with the capability of a probe to publish data on many external services, and \emph{simultaneous data publishing} concerns with the capability of a probe to publish the data on multiple services simultaneously.

\approach is the only class of probes supporting all these capabilities. Deployment of additional code is only supported by JCatascopia agents and Zabbix agents, but in both cases extending the set of the supported publishers is not possible.

%with some limitations. In fact, none of the two allows extending the set of supported publishers.

Runtime adaptation of the behaviors is very unique to \approach. While few other solutions support reconfiguration through APIs (JCatascopia agent, Datadog agent, and Zabbix agent) and within-probe analysis (JCatascopia agent and Adaptive Fogmon Follower), none of the other approaches provide self-adaptive behaviors. The only exception is Adaptive Fogmon Followers which can be configured to self-adapt the sampling rate of an indicator or to modify the set of collected indicators according to an internal analysis of the trends.
Finally, while some approaches can flexibly support multiple data ingestion services, none of the other probing systems can simultaneously interact with multiple data ingestion services.

\subsection{Scenario-Based Comparison}

\begin{table*}[!ht]
\centering
\caption{Scenario-Based Comparison}
\label{tab:scenario-based-comparison}
\resizebox{\textwidth}{!}{%
\begin{tabular}{@{}lccccccc@{}}
\toprule
\multicolumn{1}{c}{\multirow{2}{*}{\textbf{Scenario}}} &
  \multicolumn{7}{c}{\textbf{Monitoring Solutions}} \\ \cmidrule(l){2-8} 
\multicolumn{1}{c}{} &
  \textit{\approach} &
  \textit{JCatascopia~\cite{trihinas2014jacatascopia}} &
  \textit{Adaptive Fogmon~\cite{colombo2022towards}} &
  \textit{ElasticStack~\cite{elasticstack}} &
  \textit{Prometheus~\cite{prometheus}} &
  \textit{Datadog~\cite{datadog}} &
  \textit{Zabbix~\cite{zabbix}} \\ \cmidrule(r){1-1}
  
\begin{tabular}[l]{@{}l@{}}Monitoring a new data center spun up\\to deliver services to a new user base\end{tabular} & API-driven & API-driven & manual & manual & manual & \begin{tabular}[c]{@{}c@{}}partially\\API-driven\end{tabular} & \begin{tabular}[c]{@{}c@{}}partially\\API-driven\end{tabular} \\ \midrule[0.1pt]

% Streaming indicators to a new party & API-driven & manual & manual & manual & manual & manual & API-driven \\ \midrule[0.1pt]

\begin{tabular}[l]{@{}l@{}}Adapting the sampling logic\\in response to indicator trends\end{tabular} & self-adaptive & \begin{tabular}[c]{@{}c@{}}partially\\self-adaptive\end{tabular} & \begin{tabular}[c]{@{}c@{}}partially\\self-adaptive\end{tabular} & manual & manual & \begin{tabular}[c]{@{}c@{}}partially\\API-driven\end{tabular} & \begin{tabular}[c]{@{}c@{}}partially\\API-driven\end{tabular} \\ \bottomrule
\end{tabular}%
}
\end{table*}

We discuss two realistic monitoring scenarios that capture key needs in service and infrastructure monitoring, providing further evidence of the need for self-adaptiveness.

\paragraph{Monitoring a new data center spun up to deliver services to a new user base}
%\textbf{Motivation}: 
The spin-up of a new data center requires monitoring the satisfaction of a set of Service Level Objectives (SLOs) to verify the correct rollout of the system, which in turn may require both reconfiguring the monitoring system to collect new indicators and updating the sampling rate of the already collected indicators on other data centers.
%To satisfy the new requirements, it is necessary to  (1) reconfigure the monitoring system to collect the new set of indicators, and (2) update the sampling rate of the already collected indicators.

%Non adaptive monitoring solutions \textbf{TBD}
%\approach ...\textbf{TBD}
 
%\textbf{Expected Outcome}: The set of monitored indicators is extended with the new ones and the configurations are updated as required by the SLOs.
%\textbf{Example}: Lily works for a cloud provider that offers PaaS services across the world. Her company is rolling out a new data center to deliver their custom container orchestrator to a new user base. Also, they introduced a feature to support a novel network stack for premium customers in their container orchestrator. The roll out process requires updates about the SLOs in terms of both the required indicators and the frequency of their collection, in order to detect potential Service Level Agreement (SLA) violations and verify the correctness of the new features. In particular, it is required to monitor the time to provision and remove the new container network stack, and it is required to update the sampling rate of the scheduling-related indicators because several premium customers reported undetected SLA violations about the scheduling time in other data centers. To satisfy the new requirements, Lily has to (1) reconfigure the monitoring system to collect the new set of indicators, and (2) update the sampling rate of the already collected indicators.

\paragraph{Adapting the sampling logic in response to indicator trends}
%\textbf{Motivation}: 
Due to the need to promptly detect anomalies to avoid potential system failures or SLA violations, it is required to dynamically adapt both the collection and sampling logic of an indicator according to the analysis of its trend.

%Non adaptive monitoring solutions \textbf{TBD}
%\approach ...\textbf{TBD}

%\textbf{Expected Outcome}: The monitoring systems is dynamically reconfigured to change the sampling rate and/or the collection logic of some indicators in response to the indicator's trends.
%\textbf{Example}: Tracy is a software engineer working for a network operator providing Network-as-a-Service solutions. She is responsible for a new product where they need to monitor the bandwidth for both optimizing the product and checking customer's SLAs. Since passive measurements are less intrusive, but are accurate only within certain ranges~\cite{forti2021lightweight}, Tracy is in charge of introducing an approach to (1) switch collection logic of the bandwidth based on the current bandwidth range (e.g., switching from passive to active bandwidth monitoring), and (2) reconfigure the sampling rate coherently.

%\smallskip
Table~\ref{tab:scenario-based-comparison} summarizes how each scenario is supported by the monitoring solutions with the following labels:
%\begin{itemize}
\textit{manual}, the scenario can be completed by doing  manual operations; % (e.g., updating configuration files, restarting probes, deploying new components);
\textit{partially API-driven}, the scenario can be partially completed using an API, but some operations still require manual intervention;
\textit{API-driven}, the scenario can be entirely completed by interacting with APIs;
\textit{partially self-adaptive}, the scenario can be addressed with self-adaptive behaviors, but some API interactions or manual operations are still needed;
\textit{self-adaptive}, the scenarios can be completed using self-adaptive behaviors.
%\end{itemize}

The solution providing the highest level of automation is \approach, which provides probes that can be entirely reconfigured through APIs or configured with self-adaptive behaviors. The closest solution in terms of automation based on API is Zabbix, which provides a full spectrum of API-driven automation, but does not provide self-adaptive capabilities. Some self-adaptive capabilities are present in JCatascopia and Adaptive Fogmon. However, self-adaption is limited to some specific aspects, such as specific settings (e.g., sampling rate) or part of the workflow (e.g., adaptive data analysis only), and API-driven reconfiguration is partially or not present at all. The rest of the solutions provide limited reconfiguration capabilities and no self-adaptation capabilities.

%% file: sections/04-related-work.tex
 %!TeX root = main.tex
 
\section{Related Work}
\label{sec:related-work}
 %Adaptive monitoring  has been investigated in various ways, from simple monitoring system reconfiguration to more complex component re-compositions~\cite{zavala2019adaptive}. This section focuses on related work concerning probe-level adaptivity and reconfiguration.

% In the domain of adaptive monitoring and dynamic configuration in cloud environments, various research efforts and solutions have emerged, addressing the challenges posed by evolving  requirements. We categorize related work into two groups: those focusing on Adaptive Probing, and those concentrating on Adaptive Monitoring.

% \paragraph{Adaptive Probing}
%Several approaches leverage the concept of adaptive probes, offering tailored solutions to dynamically adjust monitoring probes based on changing conditions. These works contribute to the evolving domain of adaptive monitoring, each one emphasizing distinct aspects of probe adaptability, such as configuration, life-cycle management and sampling.

Adaptive monitoring is an approach investigated in many domains to respond to continuously changing conditions~\cite{zavala2019adaptive}. For example, the probe framework~\cite{estrada2015dynamic} uses adaptive probes for the dynamic creation of custom detectors, allowing their run-time insertion and removal. 
The Netconf-based approach~\cite{munz2006using} leverages adaptive monitoring probes for anomaly and attack detection scenarios. %Overcoming SNMP and CLI limitations, it provides a device-independent XML configuration data model, covering common parameters for flow measurement, aggregation, packet sampling, and data export. 
JCatascopia~\cite{trihinas2014jacatascopia} introduces adaptive probes to enable dynamic adjustments based on the detected workload variance. %These probes possess various capabilities, adding custom metric value checks, configuring collecting periods and providing filtering. 
Tundo\etal~\cite{tundo2023automated} proposed a framework for automated probe life-cycle management within the context of Monitoring-as-a-Service for cloud systems. %This framework allows dynamic deployment and undeployment of probes, simplifying adaptation to new monitoring requirements. 
A multitude of studies, including AdapPF~\cite{huang2023adappf} and the approach by Mertz and Nunes~\cite{mertz2023software}, delve into adaptive strategies for sampling metrics in dynamic environments. %AdapPF dynamically adjusts the collection frequency of monitoring data for geo-distributed cluster federations, aiming to achieve precise monitoring data for accurate application scheduling with lower cross-cluster network traffic. Similarly, Mertz and Nunes propose an approach that dynamically adjusts the sampling rate of execution traces based on the current workload of the system to reduce monitoring overhead.
 In contrast to these works, \approach introduces an architecture for the design of probe-level self-adaptive behaviors. % \approach offers operators the possibility to engineer self-adaptive behaviors at a fine-grained level of control. While other solutions may focus on specific aspects of adaptive probing, \approach provides a framework for engineering adaptive behaviors, empowering operators with dynamic reconfigurability at probe-level.

%% file: sections/05-conclusions.tex
\section{Conclusions}
\label{sec:conclusions}
The evolving nature of distributed systems demands for reconfigurable and adaptive monitoring solutions capable of swiftly responding to dynamic changes in both the requirements and the observed system. This paper introduces \emph{ReProbe}s, an architecture for reconfigurable and adaptive probes with probe-level self-adaptive behaviors. Unlike traditional systems requiring probe deployment and undeployment, \approach offers flexibility without disruption.

%The design incorporates managers and variability points, allowing configurable self-adaptive capabilities. Managers control data transmission, collection processes, and active adaptive behaviors, ensuring adaptability to diverse run-time scenarios. Variability points provide flexibility in choosing instances for specific use cases.

% Early results indicate its potential for dynamic adjustments, addressing the challenges of modern cloud environments.
\approach stands as a significant, although preliminary, step towards efficient and responsive monitoring in the era of complex distributed architectures. Our future work concerns with the quantitative experimentation of the proposed approach, beyond the qualitative evaluation presented in this paper.